\documentclass[a4paper,10pt]{article} 
\input epsf.tex

\newcommand{\be}{\begin{equation}}
\newcommand{\ee}{\end{equation}}
\newcommand{\ba}{\begin{eqnarray}}
\newcommand{\ea}{\end{eqnarray}}
\newcommand{\ban}{\begin{eqnarray*}}
\newcommand{\ean}{\end{eqnarray*}}

\newcommand{\braket}[2]{\mbox{$ \langle #1 | #2 \rangle $}}

\newcommand{\ket}[1]{\mbox{$ | #1 \rangle $}}

\newcommand{\demi}{\frac{1}{2}}

\newcommand{\one}{\leavevmode\hbox{\small1\normalsize\kern-.33em1}}

\begin{document}

\title{\Large \sc{Four-photon correction in two-photon Bell experiments}}
{\normalsize{\author{Valerio Scarani, Hugues de Riedmatten, Ivan
Marcikic, Hugo Zbinden, Nicolas Gisin\\Group of Applied Physics,
University of Geneva\\20, rue de l'Ecole-de-M\'edecine, CH-1211
Geneva 4, Switzerland}}}
\date{\today}
\maketitle

\begin{abstract}
Correlated photons produced by spontaneous parametric
down-conversion are an essential tool for quantum communication,
especially suited for long-distance connections. To have a
reasonable count rate after all the losses in the propagation and
the filters needed to improve the coherence, it is convenient to
increase the intensity of the laser that pumps the non-linear
crystal. By doing so, however, the importance of the four-photon
component of the down-converted field increases, thus degrading
the quality of two-photon interferences. In this paper, we present
an easy derivation of this nuisance valid for any form of
entanglement generated by down-conversion, followed by a full
study of the problem for time-bin entanglement. We find that the
visibility of two-photon interferences decreases as $V=1-2\rho$,
where $\rho$ is, in usual situations, the probability per pulse of
creating a detectable photon pair. In particular, the decrease of
$V$ is independent of the coherence of the four-photon term.
Thanks to the fact that $\rho$ can be measured independently of
$V$, the experimental verification of our prediction is provided
for two different configuration of filters.
\end{abstract}

\section{Introduction}

The distribution of a pair of entangled photons to two distant
partners is the building block of quantum communication protocols
\cite{qcomm,tw}. The entangled photons are produced by parametric
down-conversion (PDC) in a non-linear crystal. As well-known, this
process creates pairs of photons at the first order; but when the
pumping intensity increases, four- and more-photon components
become important in the down-converted field \cite{mandel,walls}.
If one can post-select the number of photons, higher-photon
components of the field may turn out to be a useful resource (an
"entanglement laser", see \cite{entglaser}). In other cases
however, these higher-number components turn out to be quite a
nuisance. In particular, one is often interested in two-photon
phenomena, just think of the Bell-state measurement (BSM) that is
needed in teleportation. The presence of higher-number components
obviously degrades the quality of the two-photon interferences. In
long-distance implementations, one can hardly overcome this
nuisance by working with low pump intensities: after propagation
along several kilometers of fibers, many photons are lost because
of the losses in the fibers, and the efficiency of the detectors
at telecom wavelengths is low, typically 10\%. Moreover, if the
two photons come from different sources and have to interfere at a
beam-splitter (as is the case for the BSM), filters must be
introduced to ensure coherence. Thus, in order to have a
reasonable count rate, one has to increase the pump power
--- and this unavoidably increases the number of unwanted
higher-number components. In this paper, we address the
degradation of the visibility of two-photon interferences due to
the presence of four-photon components in the field, thus
completing the partial study provided in Ref. \cite{ivanpra}.

In Section \ref{secint}, we give an easy derivation of the topic
and the results that is valid for any form of entanglement
generated by down-conversion, under the assumption that the
four-photon component is described by two independent pairs. In
the rest of the paper, we relax that assumption: indeed, the
four-photon coherence can vary from zero (two independent pairs)
to one (state of single-mode down-conversion \cite{weinf})
according to the experimental conditions
\cite{john,hugues,hoffman}. We prove that the loss of visibility
does not depend on the coherence of the four-photon state, but
only on a parameter $\rho$ that is basically the probability of
creating a detectable pair. For this full study, we shall focus on
{\em time-bin entanglement}, a form of entanglement that is more
robust than polarization for long-distance applications in optical
fibers. Visibilities large enough to allow the violation of Bell's
inequalities for two photon \cite{ivanpra,otherbell}, quantum
cryptography \cite{crypto}, and long-distance teleportation
\cite{teleport} have been demonstrated in the recent years for
this form of entanglement. For time-bin entanglement, the present
study requires the multimode formalism, introduced in Section
\ref{secgen}. In fact, as the name suggests, a time-bin qubit is a
coherent superposition of two orthogonal possibilities, the photon
being at a given time $t=0$ (first time-bin) or at a later time
$t=\tau$ (second time-bin). Separate time-bins must be created by
a pump field consisting of separate pulses: the finite temporal
size (thence, the non-monochromaticity) of the pump pulses and the
down-converted photons is a necessary feature of time-bin qubits.

In Section \ref{seccal}, we describe a setup that is used for
measure the parameter $\rho$. In Section \ref{secfran}, we
introduce the setup for measuring time-bin entanglement (a Franson
interferometer with a suitable source) and derive our main
prediction, namely the decrease of visibility due to the presence
of four-photon components in the state. In Section \ref{secexp} we
describe the experimental verification of our predictions. Section
\ref{seccon} is a conclusion. For readability, the technicalities
of the formalism used in Sections \ref{seccal} and \ref{secfran}
are left for an Appendix. We note that the calculations of the
two-photon coincidence rate provides the first explicit
calculation of time-bin Bell experiments using the full formalism
of quantum optics.

\section{Easy derivation for incoherent four-photon component}
\label{secint}

\begin{figure}
\begin{center}
\epsfxsize=9cm \epsfbox{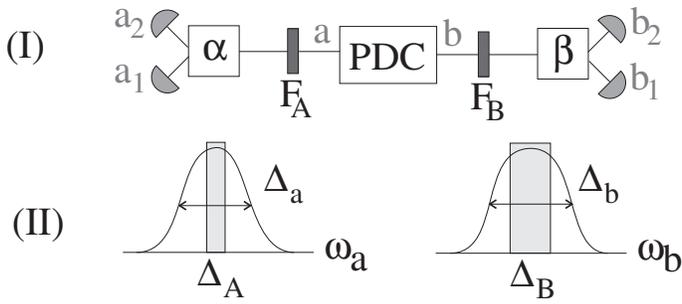} \caption{(I) Experimental
setup to measure two-photon interferences, and meaningful
parameters. Grey letters: spatial modes in the fibers; PDC:
parametric down-conversions; F: filters. (II) Spectral widths of
the down-converted photons (curves) and filters (grey shadows).
Both filters are centered in the spectrum of the down-converted
photons, and their width are $\Delta_A\leq \Delta_B$.}
\label{schemafilters}
\end{center}
\end{figure}

The purpose of this Section is to derive the main results from a
simple formalism, in order to gain intuition about the physics of
the problem. The content of this Section does not apply only to
time-bin entanglement, but to any form of entanglement obtained by
down-conversion, be it with a cw or with a pulsed pump laser. The
probabilities that we are going to introduce in this Section are
"per detection window". In the case of a cw pump, this means "per
time resolution of the detector"; in the case of a pulsed pump,
this means "per pump pulse" ("per qubit", in the language of
time-bin entanglement \cite{noteper}).

The calculation is possible in simple terms if we neglect the
coherence of the four-photon term, and assume that when four
photons are produced, they form two independent pairs. The process
of creation of independent items obeys the Poissonian statistics:
if $P_{2c}$ is the probability of creating a pair, we have
$P_{4c}\simeq \demi P_{2c}^2$. For the setup, we refer to Fig.
\ref{schemafilters}. We define $\Delta_{a,b}$ as the spectral
width of the photons in mode $a$, resp. $b$, after
down-conversion, that is, before the filters; the spectral width
of the pump is denoted $\Delta_p$. As for the filters, we suppose
that they are centered in the spectrum of the down-converted
photons, and that they satisfy $\Delta_{A,B}\leq\Delta_{a,b}$ to
avoid trivialities; furthermore, we suppose
$\Delta_{A,B}>>\Delta_{p}$, so that twin photons certainly pass
both filters, and $\Delta_B\geq \Delta_A$. Let's follow the two-
and the four-photon component through the setup, until the
coincidence detection in modes $a_1$ and $b_1$.

{\em Two-photon component.} To have a detection, both photons must
pass the filters; because of the correlation in energy, if photon
$a$ passes through $F_A$ (that happens with probability $\sim
\Delta_A/\Delta_a$), then certainly photon $b$ will pass through
$F_B$, because this filter is larger and the photons are
correlated in energy. The photons are twins, therefore they
interfere. Consequently, the detection rate due to two-photon
components is (up to multiplicative factors) \ba R_2 &=&
P_{2c}\,\frac{\Delta_A}{\Delta_a}\,\demi\big[1+
\cos(\alpha+\beta)\big]\,. \ea

{\em Four-photon component.} Once four photons have been produced,
four two-photon coincidence events are possible: two events in
which we detect photons belonging to the same pair, and two events
in which we detect photons belonging to different pairs. The first
case is similar to the case of two-photons. In the second case,
however, the fact that photon $a$ passes its filter does not
guarantee at all that photon $b$ will do it as well; and of
course, no interference will take place. All in all, \ba R_4 &=&
P_{4c}\,\left\{2\frac{\Delta_A}{\Delta_a}\,\demi\big[1+
\cos(\alpha+\beta)\big] \,+\, 2 \frac{\Delta_A}{\Delta_a}
\frac{\Delta_B}{\Delta_b}\,\demi \right\}\,. \ea The total count
rate is therefore $R_2+R_4 = \bar{R}\,\demi\big[1+
V\cos(\alpha+\beta)\big]$ where $\bar{R}\simeq
P_{2c}\frac{\Delta_A}{\Delta_a}$ and where the visibility $V$ is
\ba
V&=&\frac{1}{1+\frac{P_{2c}}{1+P_{2c}}\frac{\Delta_B}{\Delta_b}}\,
=\, 1\,-\,P_{2c}\frac{\Delta_B}{\Delta_b}\,+\,O(P_{2c}^2)\,.
\label{visint}\ea Recall that $\Delta_B$ is singled out by the
relation $\Delta_B\geq \Delta_A$. As expected, $V$ decreases if
$P_{2c}$ (proportional to the pump power) is increased. Note also
that $\frac{\Delta_B}{\Delta_b}\leq 1$: for a given pump power,
the visibility increases if filters are in place. This is
intuitive, considering the emission of two pairs: conditioned to
the fact that photon in mode $a$ has passed the filter $F_A$, a
photon passing $F_B$ is more likely to be its twin (whose
frequency {\em must} lie within the filter) than an uncorrelated
photon (whose frequency may lie everywhere in the spectrum).
Finally, if only one filter is in place, then
$\frac{\Delta_B}{\Delta_b}=1$ and we recover the discussion
presented in Ref. \cite{ivanpra}.

However, we are not really interested in fixing the pump power:
rather, we'd like to fix the coincidence rate at the detection
$\bar{R}$. Obviously, this means that if we narrow the filters, we
must increase the pump power in order to keep the coincidence rate
constant. Strictly speaking, the quantity
$P_{2c}\frac{\Delta_B}{\Delta_b}$ is the probability per qubit of
creating a photon pair such that the photon in mode $b$ passes
through the (larger) filter. However,
$\frac{\Delta_A}{\Delta_a}\simeq \frac{\Delta_B}{\Delta_b}$ holds
in magnitude for typical down-conversion processes and filters;
consequently, $P_{2c}\frac{\Delta_B}{\Delta_b}\simeq \bar{R}$ is
an estimate of the probability of creating a detectable pair.

The results of this Section are based on the assumption that the
four-photon state is always described by two independent pairs.
Note that this assumption is certainly good in the case of cw
pump, because the time resolution of the detector is much larger
than the coherence time of the down-converted photons. The
assumption is more questionable in the case of a pulsed pump. The
rest of the paper shows, focusing specifically on time-bin
entanglement, that the degradation of visibility (\ref{visint}) is
actually independent of the coherence of the four-photon term.

\section{General approach}
\label{secgen}

\subsection{The state out of down-conversion}

The formalism to describe multimode down-conversion was introduced
in Refs \cite{keller,grice} for the two-photon component, and
extended to the four-photon component for type-I down-conversion
in \cite{wang}. We have applied this formalism to our case in Ref.
\cite{hugues}; we summarize here the main notations and results.

The pump field is assumed to be classical, composed of two
identical but delayed pulses: $P(t)=\sqrt{I_p}\,\big(p(t)
+p(t+\tau)\big)$, so in Fourier space \ba
\tilde{P}(\omega)&=&\sqrt{I_p}\,\tilde{p}(\omega)\,\left(1+e^{i\omega\tau}\right)\,.
\label{pulse} \ea We use colinear type-I down-conversion in a
non-degenerate regime $\omega_s\neq \omega_i$; therefore, the
signal and the idler photons can be coupled into different spatial
modes $a$ and $b$ using a wavelength division multiplexer (WDM).
The phase-matching function is written $\Phi(\omega_a,\omega_b)$;
we don't need its explicit form in what follows. For convenience
we define the following notations: \ba
\Phi(x,y)\,\tilde{p}(x+y)\,(1+e^{i(x+y)\tau})&\equiv\,g(x,y)\,(1+e^{i(x+y)\tau})&\equiv\,G(x,y)\,.
\ea The state produced by the down-conversion in the crystal reads
\ba \ket{\Psi}&=&i\sqrt{I}\,{\cal A}^{\dagger}\ket{vac} \,+\,
\frac{I}{2}\,\big({\cal
A}^{\dagger}\big)^2\ket{vac}\,+\,O(I^{3/2})\ea where $I$ is
proportional to the intensity $I_p$ of the pump, and \ba {\cal
A}^{\dagger}&=&\int d\omega_ad\omega_b
G(\omega_a,\omega_b)a^{\dagger}(\omega_a)b^{\dagger}(\omega_b)\,.
\ea

\subsection{Detection: generalities}

We have just given the state $\ket{\Psi}$ created by
down-conversion. This state evolves through the setup (in our
case, a linear optics one so that the number of photons is
conserved) according to $\ket{\Psi}\rightarrow\ket{\hat{\Psi}}$,
then two-photon coincidences are recorded. Here we introduce the
general scheme for this detection. Let's write $a_1$ and $b_1$ the
spatial modes on which one looks for coincidences; since no
ambiguity is possibly, we write $a_1$ and $b_1$ also the
corresponding annihilation operators. We look at detector on mode
$a_1$ at time $T_A\pm\Delta T$, where $\Delta T$ is the time
resolution of the detectors; similarly for detection on mode
$b_1$. The coincidence rate reads \ba
R(T_A,T_B)&=&\eta_A\eta_B\,\int_{T_A-\Delta T}^{T_A+\Delta T}dt_A
\int_{T_B-\Delta T}^{T_B+\Delta T}dt_B\,|| E_{a_1}^{(+)}(t_A)\,
E_{b_1}^{(+)}(t_B)\,\ket{\hat{\Psi}}||^2 \label{detgen}\,.\ea In
this formula, $\eta_{A,B}$ are constant factors \cite{note0} that
will be omitted in all that follows; the positive part of the
electric field on mode $a_1$ is defined as \ba E_{a_1}^{(+)}(t)&=&
\,\int d\nu f_A(\nu)\,e^{-i\nu t}\,a_1(\nu)\,. \ea with $f_A(\nu)$
is a real function describing a filter in mode $a_1$, the
transmission of the filter being $F_A(\nu)=f_A(\nu)^2$. The
definition of $E_{b_1}^{(+)}(t)$ is analogous. We choose the
origin of times in order to remove the free propagation from the
crystal to the detectors. Therefore, the first time-bin at the
detection is given by $t_j=0$, the second time-bin by $t_j=\tau$
and so on.

Actually, formula (\ref{detgen}) for detection is exact for
proportional counters, in which the probability of detection is
the intensity of the field. For photon counting with a detector of
quantum efficiency $\eta$, the probability of the detector firing,
given that $n$ photons imping on it, is not $n\eta$ (proportional
to the intensity) but $(1-(1-\eta)^n)$. Now, for the wavelengths
that we consider, the quantum efficiency is $\eta\approx 0.1$;
moreover, the mean number of photons that imping on a detector is
much smaller than 1 because of the losses in the fibers and in the
coupling; finally, in our formalism we restrict to the four-photon
term, so that at most two photons can imping on the detector. All
in all, the approximation $(1-(1-\eta)^2)\simeq 2\eta$ holds and
we can indeed use (\ref{detgen}) to compute the coincidence rate.

\subsection{Important parameters}
\label{subparam}

As we said in the introduction, we shall postpone the detailed
calculations to the Appendix. All the results of the Sections
\ref{seccal} and \ref{secfran} can be formulated using the
following parameters: writing
$d\underline{\omega}=d\omega_ad\omega_b$, \ba J&=&\int
d\underline{\omega}\, \left|g(\omega_a,\omega_b)\right|^2\,,
\label{defjnew}\\
J_A&=& \int d\underline{\omega}\,F_A(\omega_a)\,
\left|g(\omega_a,\omega_b)\right|^2\\
J_B&=& \int d\underline{\omega}\,F_B(\omega_b)\,
\left|g(\omega_a,\omega_b)\right|^2\\
J_{AB}&=&\int d\underline{\omega}\,F_A(\omega_a)F_B(\omega_b)\,
\left|g(\omega_a,\omega_b)\right|^2\,, \label{defjab}\\
J_4&=&\int d\underline{\omega}
d\underline{\omega}'\,F_A(\omega_a)F_B(\omega_b)\,
\left[g^*(\omega_a,\omega_b)g^*(\omega_a',\omega_b')g(\omega_a,\omega_b')
g(\omega_a',\omega_b)\,+\,c.c.\right]\,, \label{defl} \ea The
first four numbers can be given an intuitive meaning. In fact, up
to multiplicative factors: $J$ is the probability of producing two
photons in one pump pulse, irrespective of whether they will pass
the filter or not; $J_A$ and $J_B$ are the probabilities of
producing two photons in one pump pulse, and that the photon in
mode $a$ (resp. $b$) passes through the filter; $J_{AB}$ is the
probability of producing two photons in one pump pulse and both
photons pass the filter. The interpretation of $J_4$ is somehow
more involved: it is a coherence term, due to the fact that the
four photon state cannot be described as two independent pairs
\cite{hugues}.

Obviously, $J_{AB}=J_A$ if no filter is applied on B. But
$J_{AB}=J_A$ holds to a very good approximation also if
$\Delta_A<\Delta_B$, where $\Delta_X$ is the width of filter
$F_X$, provided that both filters are larger than the spectral
width of the pump $\Delta_p$ (as we supposed in Section
\ref{secint}, and as will be the case in the experiment). In fact,
in this case, detection of a photon in filter A automatically
ensures that its twin photon has a frequency within the range of
filter B, which means $F_B(\omega_b)=1$ for all $\omega_b$
compatible with the phase-matching condition.

\section{A calibration setup}
\label{seccal}

\begin{figure}
\begin{center}
\epsfxsize=9cm \epsfbox{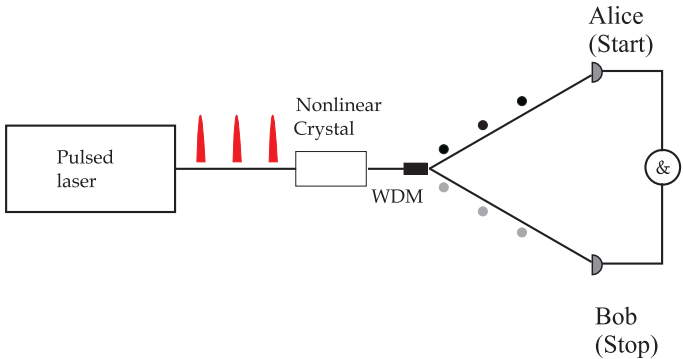} \caption{Schematic of the
setup used to measure the parameter $\rho$.} \label{setupbaby}
\end{center}
\end{figure}

Before describing the measurement of two-photon interferences
(next Section) we present an experimental setup that allows to
measure the probability $\rho$ of creating a detectable pair in a
simple way. This setup (see Fig. \ref{setupbaby}) has been
presented in detail in section IV of Ref. \cite{ivanpra}. We give
here a brief description. A Fourier-transform-limited pulsed laser
is used to create non-degenerate photon pairs at telecommunication
wavelengths (1310 and 1550 nm) by parametric down-conversion in a
non linear-crystal. The two photons are separated
deterministically using a wavelength-division multiplexer (WDM)
and each photon is detected by single-photon counters (avalanche
photodiodes). The signal from the two detectors are then sent to a
Time-to-Digital converter, which is used to determine the
histogram of the differences in the time of arrival of the twin
photons.

We apply our formalism to this setup. For the detection, since
there is no evolution but the free propagation, we have simply
$a_1=a$ and $b_1=b$. For the preparation, at first sight it seems
that our formalism should be modified: we are dealing with a train
of $N$ pulses instead of only two pulses, so
$\left(1+e^{i\omega\tau}\right)$ should be replaced with
$\sum_{k=0}^{N-1}e^{i\omega\,k\tau}$ in formula (\ref{pulse}).
However, a closer look shows that we can do the calculation
without any change. In fact, in this particular setup there is no
interference: then, $R_C$ is simply the sum of the coincidence
rates obtained when the two photons arrive at the same time, while
$R_L$ is the sum of the coincidence rates obtained when photon in
mode $a$ arrives a time $\tau$ later than the photon in mode $b$.
Since moreover $R(k\tau,k\tau)=R(0,0)$ and
$R((k+1)\tau,k\tau)=R(\tau,0)$ for all $k$, we obtain
$R_C=N\,R(0,0)$ and $R_C=(N-1)\,R(\tau,0)$, so we can focus on
only two successive pulses. By the way, $R(0,0)$ is proportional
to the probability per pulse of creating one detectable pair (a
pair that will pass the filters).

The calculation is given in the Appendix, and the results are
$R(0,0)\,=\,I\,J_{AB}+O(I^2)$ and $R(\tau,0)\,=\,I^2\,J_AJ_B$,
that are indeed what one expects because of the meaning of the
$J$'s (subsection \ref{subparam}). Therefore, in the limit of
large $N$, the ratio $\rho$ between the integrals of the side peak
and the central peak is \ba \rho&=&\,I\,\frac{J_AJ_B}{J_{AB}}\,.
\label{ppair}\ea In most cases, $\rho$ has a simple
interpretation. In fact, whenever condition
$\Delta_p<<\Delta_A<\Delta_B$ holds, we have seen above that
$J_{AB}=J_A$ and consequently $\rho=IJ_B$ is the probability per
pulse of creating a pair such that the photon that meets the
largest filter will pass it. In particular, if there is no filter
on mode $b$, $\rho$ is the probability per pulse of creating a
pair, as noticed in the Appendix of \cite{ivanpra}. That
derivation shares with the present one the hypothesis of small
detector efficiency, but is otherwise rather different: in our
previous paper, we supposed that a $2N$-photon state is actually
$N$ independent pairs; here, we limit ourselves to 2 and 4
photons, but derive the result without any assumption about the
coherence of the 4-photon state.

Moreover, as argued in Section \ref{secint}, since $J_A\simeq J_B$
normally holds, at least in magnitude, then $\rho\simeq
IJ_A=R(0,0)$ is an estimate of the probability per pulse of
creating a detectable pair.

\section{The Franson interferometer}
\label{secfran}

\subsection{Description of the setup}

\begin{figure}
\begin{center}
\epsfxsize=9cm \epsfbox{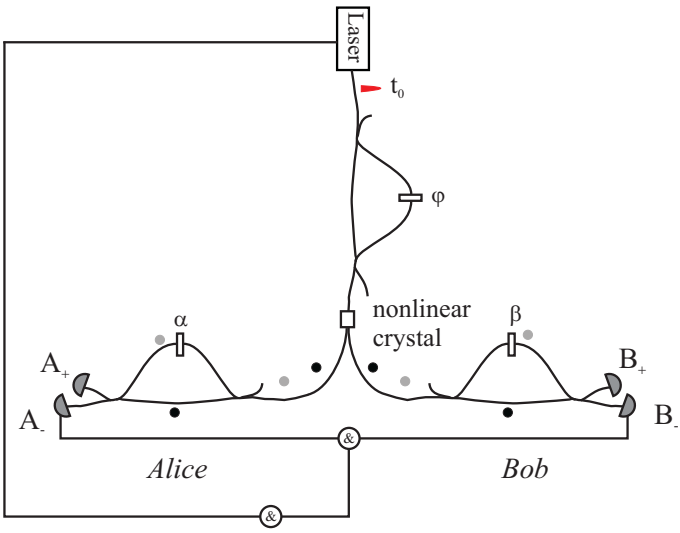} \caption{Schematic of the
setup used to measure two photon quantum interference with
time-bin entangled qubits. In addition to the two-photon
coincidence, a coincidence with the pump laser provides the origin
of time needed to define the three time-bins.} \label{setupb}
\end{center}
\end{figure}

We turn now to the main setup, which is the interferometer that
allows the analysis of time-bin entanglement (see
Fig.\ref{setupb}). This is essentially the interferometer proposed
by Franson to study energy-time entanglement \cite{franson},
completed with an unbalanced interferometer before the crystal
(the pump interferometer). A laser pulse is first split in two in
this interferometer. At its exit, we have two laser pulses with a
fixed phase difference separated by a time $\tau$ corresponding to
the path length difference between the long and the short arm of
the interferometer. In the non linear crystal, we therefore create
a photon pair in a coherent superposition of two time-bins.

After the crystal, the photons are separated with the WDM and each
sent to a fiber interferometer in order to make a two photon
interference experiment.

\subsection{Evolution}

The evolution of modes $a$ and $b$ in each arm of the
interferometer is given by the following expressions: \ba
a^{\dagger}(\omega)&\longrightarrow &
\hat{a}^{\dagger}(\omega)\,=\,S(\omega,\alpha)\,a_1^{\dagger}(\omega)
+ C(\omega,\alpha)\,a_2^{\dagger}(\omega)\\
b^{\dagger}(\omega)&\longrightarrow &
\hat{b}^{\dagger}(\omega)\,=\,S(\omega,\beta)\,b_1^{\dagger}(\omega)
+ C(\omega,\beta)\,b_2^{\dagger}(\omega) \ea with \cite{notenew}
\ba
S(\omega,\theta)\,=\,\frac{1-e^{i(\omega\tau+\theta)}}{2}\;&,&\;
C(\omega,\theta)\,=\,i\,\frac{1+e^{i(\omega\tau+\theta)}}{2}\,.
\label{sin} \ea The evolved state $\ket{\hat{\Psi}}$ is obtained
by inserting the evolved operators $\hat{a}^{\dagger}$ and
$\hat{b}^{\dagger}$ into $\ket{\Psi}$.

In these formulae, we have already supposed that the analyzing
interferometers are identical to the pump interferometer. Thus,
three time-bins are defined by the setup. The first time-bin,
$t=0$, corresponds to the time of arrival of photons produced by
the first pump pulse and not delayed. The second or intermediate
time-bin, $t=\tau$, corresponds to the time of arrival, either of
photons produced by the first pump pulse and delayed, or of
photons produced by the second pump pulse and not delayed. The
third time-bin, $t=2\tau$, corresponds to the time of arrival of
photons produced by the second pump pulse and delayed.
Interferences will only be seen when both photons arrive at
$t=\tau$, because only in this case two indistinguishable
alternatives are available.

\subsection{Two-photon interferences}

We study the detection for modes $a_1$ and $b_1$; all the other
cases can be treated in the same way. The coincidence rate
$R(T_A,T_B)$ is the sum of two terms corresponding respectively to
the two-photon and the four-photon terms: \ba
R_2(T_A,T_B)&=&I\,\int dt_Adt_B || E_{a_1}^{(+)}(t_A)\,
E_{b_1}^{(+)}(t_B)\,\hat{\cal A}^{\dagger}\ket{vac}||^2\,, \label{r2abstr} \\
R_4(T_A,T_B) &=&\frac{I^2}{4}\,\int dt_Adt_B ||
E_{a_1}^{(+)}(t_A)\, E_{b_1}^{(+)}(t_B)\,\big(\hat{\cal
A}^{\dagger}\big)^2\ket{vac}||^2\,. \label{r4abstr}\ea Indeed, the
two- and the four-photon states do not interfere (in principle,
one could insert a non-destructive measurement of the number of
photons just after the crystal, and this would not modify the rest
of the experiment).

The calculation is presented in the Appendix. As said above,
interferences will appear only in the intermediate time-bin
$T_A=T_B=\tau$, in which case one finds \cite{note3}: \ba
R_2(\tau,\tau)&=&I\,J_{AB}\,\big(1+\cos(\alpha+\beta)\big)\,,\label{r2fin}\\
R_4(\tau,\tau)&=&I^2\,\Big[\big(2J_{AB}J\,+\,J_4)\,
\big(1+\cos(\alpha+\beta)\big)\,+\,2J_AJ_B\Big]\,.
\label{r4fin}\ea The result for $R_2(\tau,\tau)$ is the expected
one: one pair is produced, it passes the filters, and since it is
in a superposition of being in both pulses it gives rise to
full-visibility interferences. In the formula for
$R_4(\tau,\tau)$, two contributions are also expected from the
intuitive view of the four-photon state as two independent pairs:
(i) the term containing $J_{AB}J$ means that two pairs are
created, the photons of the same pair are detected and therefore
one has full visibility; (ii) the term containing $J_{A}J_B$ means
that two pairs are created, the photons of the different pair are
detected and therefore they don't show any interference. The
remarkable feature is the {\em position} of the correction due to
the coherence in the four-photon term, $J_4$: it contributes to a
full-visibility interference as well. This couldn't have been
guessed without the full calculation.

Summing (\ref{r2fin}) and (\ref{r4fin}), the total two-photon
coincidence rate in the intermediate time-bin reads \ba
R(\tau,\tau)&=&R_2(\tau,\tau)+ R_4(\tau,\tau)\,=
\,\bar{R}\,\big[1\,+\,V\cos(\alpha+\beta)\big] \ea where the
average count rate $\bar{R}$ is given by \cite{note3} \ba \bar{R}=
IJ_{AB}\,+\, O(I^2) \label{average}\ea and the visibility $V$ is
given by $V =
\frac{1+I(J+J_4/J_{AB})}{1+I(J+(J_4+J_AJ_B)/J_{AB})}$. Now, the
terms $O(I^2)$ in the visibility are meaningless, because the
six-photon term that we neglected completely contributes to the
same order; so we have to keep only the first-order development of
$V$ in $I$, that leads to the remarkable relation \ba V\,\simeq\,
1\,-\,I\,\frac{2J_AJ_B}{J_{AB}}\,=\,1\,-\,2\rho \label{visi1} \ea
where $\rho$ is exactly the same as defined in (\ref{ppair}). As
announced in Section \ref{secint}, $J_4$ drops out of the
visibility: to the leading order in $I$, the loss of visibility is
independent of the coherence of the four-photon state.

Since $\rho$ is basically the probability per pulse (so that
$2\rho$ is the probability per qubit \cite{noteper}) of creating a
detectable pair, it defines the detection rate up to
multiplicative factors. Relation (\ref{visi1}) therefore says
that, if we fix a detection rate, we shall find a given
visibility, no matter whether the rate was obtained by pumping
weakly and putting no filters, or by pumping strongly and putting
narrow filters. This is a positive feature: filters, while being
useful to improve the coherence whenever this is required, do not
degrade the visibility.

As described in section \ref{seccal}, $\rho$ can be measured
independently, the relation (\ref{visi1}) can be experimentally
tested. This is the object of the next Section.

\section{Experimental verification}
\label{secexp}

In this section, we present an experimental verification of Eq.
(\ref{visi1}). Two-photon interference fringes are recorded for
different value of $\rho$, corresponding to different values of
pump power, with the Franson setup described in the previous
Section. Let us remind the reader that the down-converted photons
are at the two telecom wavelengths, 1310 nm and 1550 nm. The
measurement are reported for two different filters configurations;
in both cases, the larger filter is on the photons at 1550nm, so
this is "mode $b$".

\begin{figure}
\begin{center}
\epsfxsize=9cm \epsfbox{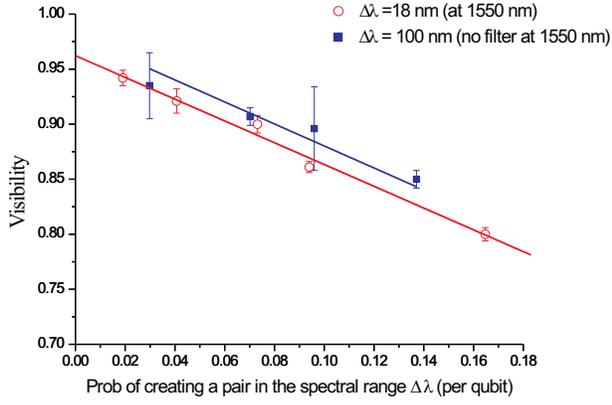} \caption{Visibility as a
function of $2\rho$ for two different filtering configurations.
Full squares are experimental points with a 40 nm filter at 1310
nm and no filter at 1550 nm (data taken from \cite{ivanpra}). Open
circles are experimental points with a 10 nm filter at 1310 nm and
a 18 nm filter at 1550 nm. The solid curves are straight line with
a slope $-1$, according to Eq. (\ref{visi1}).} \label{results}
\end{center}
\end{figure}

In the first configuration, only the photon at 1310 nm is filtered
with 40 nm FWHM. These data are taken from \cite{ivanpra}. In the
second configuration, both photon are filtered. The photon at 1310
nm is filtered with 10 nm FWHM, while the photon at 1550 nm is
filtered with 18 nm FWHM. The coefficient $\rho$ is measured using
the side peaks method explained in Section \ref{seccal}. The
visibility for the two experimental configurations is plotted as a
function of $2\rho$ in Fig. \ref{results}. The error bars on the
experimental points represent the accuracy of the fit of the
recorded interference patterns with a sine law \cite{note1}. The
two solid lines are straight lines with slope $-1$, according to
Eq. (\ref{visi1}); the small shift between the two curves is due
to the fact that the maximal visibility was not the same for both
experiments and was left free as a fitting parameter. We observe a
good agreement between theory and experiment. These results
confirm that the loss of visibility due to four-photon events is
directly related to $\rho$, regardless of the filtering that is
applied on the photons and regardless of the coherence of the
four-photon component \cite{hugues}. This is therefore a general
result very useful to estimate the effect of multi-pair creation
in an experiment in a very simple way.

\section{Conclusion}
\label{seccon}

In summary, we have found a quantitative prediction for the loss
of two-photon interference visibility due to the presence of a
four-photon component in the down-converted field. The loss of
visibility (\ref{visi1}) is determined by the parameter $\rho$
(\ref{ppair}), that is close to the probability of creating a
detectable pair. This parameter can be measured independently,
thus allowing a direct experimental verification of our
prediction. While the full calculation was worked out for time-bin
entanglement, we have presented in Section \ref{secint} a
simplified derivation that gives the same result and applies to
any form of entanglement generated by down-conversion.

We acknowledge fruitful discussions with Antonio Ac\'{\i}n,
Christoph Simon and Wolfgang Tittel.

{\em Note added in proof.} Since this work was finished, we have
learnt of two independent papers \cite{pap1,pap2} that discussed
the loss of visibility of two-photon interferences due to the
presence of higher-photon-number components. Both calculations
concern entanglement in polarization and have been done in the
single-mode formalism: this allows to take into account the
contribution of all more-photon terms and not only of the
four-photon one. The results are compatible with ours in the
regime where they can be compared (small pump power). Consider for
instance Ref. \cite{pap2}: from their eq. (2), we see that the
probability per qubit of producing a pair is
$2\tanh^2\tau/\cosh^4\tau\approx 2\tau^2$. Then our formula
(\ref{visint}) predicts $V\approx 1-2\tau^2$ for small values of
$\tau$, which indeed fits correctly the curve of Fig. 5 of Ref.
\cite{pap2} up to $\tau\approx 0.5$.

\begin{appendix}

\section{Appendix}
\label{app}

We recall the definitions of $R_2$ and $R_4$, formulae
(\ref{r2abstr}) and (\ref{r4abstr}). Although we introduced them
only in Section \ref{secfran}, the same quantities can be defined
for the setup of Section \ref{seccal}, and in fact for any setup:
a given setup will be characterized by the relation between the
preparation modes $a$, $b$, and the detection modes $a_1$, $b_1$,
a relation encoded in the operator $\hat{\cal{A}}^{\dagger}$. In
this appendix, we start by working out more explicitly the general
formulae for $R_2$ and $R_4$; we subsequently describe the
strategy that allows a simplification of these formulae (a
strategy already introduced in Ref. \cite{hugues}), and finally
compute the explicit results announced in Sections \ref{seccal}
and \ref{secfran}.

\subsection{General formula for $R_2$}

The calculation of $R_2$ goes as follows. The commutation rules
between the input modes $a$, $b$ and the detected modes $a_1$,
$b_1$ read \ba a_1(\nu_a)\hat{a}^{\dagger}(\omega_a)&=&
\hat{a}^{\dagger}(\omega_a)a_1(\nu_a)\,+\,E(\omega_a,\alpha)\,
\delta(\omega_a-\nu_a)\,\one \\
b_1(\nu_b)\hat{b}^{\dagger}(\omega_b)&=&
\hat{b}^{\dagger}(\omega_b)b_1(\nu_b)\,+\,E(\omega_b,\beta)\,
\delta(\omega_b-\nu_b)\,\one\ea where $E(\omega,\gamma)$ is a
function that depends on the evolution undergone by the modes from
the production to the detection --- specifically,
$E(\omega,\gamma)=1$ for the calibration setup, while
$E(\omega,\gamma)=S(\omega,\gamma)$ given in (\ref{sin}) for the
Franson setup. Using these commutation relations, one finds
immediately $E_{a_1}^{(+)}(t_A)\, E_{b_1}^{(+)}(t_B)\,\hat{\cal
A}^{\dagger}\ket{vac}=c(t_A,t_B)\ket{vac}$ where we have
introduced the complex number \ba c(t_A,t_B)&\equiv & \int
d\underline{\omega}\,
f_A(\omega_a)f_B(\omega_b)\,G(\omega_a,\omega_b)\,
E(\omega_a,\alpha)E(\omega_b,\beta)\,
e^{-i(\omega_at_A+\omega_bt_B)}\,. \label{fctc}\ea Consequently,
$R_2=I\int dt_Adt_B|c(t_A,t_B)|^2$. The integration over $t_A$ and
$t_B$ can be performed before the integrals over the frequencies,
so finally \ba R_2(T_A,T_B)&=\,I\,\int d\underline{\omega}
d\underline{\omega}'&f_A(\omega_a)
f_B(\omega_b)\,g(\omega_a,\omega_b)
f_A(\omega_a')f_B(\omega_b')\, g^*(\omega_a',\omega_b')\nonumber\\
&&\times\,\left(1+e^{i(\omega_a+\omega_b)\tau}\right)
\left(1+e^{-i(\omega_a'+\omega_b')\tau}\right)\,
{\cal{E}}(\omega_a,\omega_b,\omega_a',\omega_b') \nonumber\\
&&\times
e^{-i(\omega_a-\omega_a')T_A}\,e^{-i(\omega_b-\omega_b')T_B}\,(\Delta
T)^2\,\mbox{sinc}[(\omega_a-\omega_a')\Delta T]
 \,\mbox{sinc}[(\omega_b-\omega_b')\Delta T] \label{r2tout}\ea
where we have defined the shortcut \ba
{\cal{E}}(\omega_a,\omega_b,\omega_a',\omega_b')&=&
E(\omega_a,\alpha)E(\omega_b,\beta)
E^*(\omega_a',\alpha)E^*(\omega_b',\beta)\,.  \ea In
(\ref{r2tout}), the first two lines are simply the expansion of
the $G$'s and the evolution term ${\cal{E}}$; the last line is the
result of the integration over $t_A$ and $t_B$.

\subsection{General formula for $R_4$}

The calculation of $R_4$ follows exactly the same structure as the
calculation of $R_2$, only the formulae are heavier. Using \ba
a_1(\nu)\hat{a}^{\dagger}(\omega)\hat{a}^{\dagger}(\omega') &=&
\hat{a}^{\dagger}(\omega)\hat{a}^{\dagger}(\omega')a_1(\nu)+
E(\omega,\alpha) \delta(\omega- \nu)\hat{a}^{\dagger}(\omega')+
E(\omega',\alpha)\delta(\omega'- \nu)\hat{a}^{\dagger}(\omega)\ea
and the analogous relation for mode $b$, one finds \ba
E_{a_1}^{(+)}(t_A)\, E_{b_1}^{(+)}(t_B)\,\big(\hat{\cal
A}^{\dagger}\big)^2\ket{vac}&=&\ket{AB}+\ket{A'B'}+\ket{A'B}+\ket{AB'}\equiv
2(\ket{AB}+\ket{AB'})\,.\label{epsi4} \ea We have defined \ban
\ket{AB}&=& \int d\underline{\omega} d\underline{\omega}'\,
G(\omega_a,\omega_b)G(\omega_a',\omega_b')\,
Z(\omega_a,\omega_b,\omega_a',\omega_b')\ket{vac}\ean where $Z$ is
the non-normalized two-photon creation operator \ban
Z(\omega_a,\omega_b,\omega_a',\omega_b')&=&f_A(\omega_a)
f_B(\omega_b)\, e^{-i(\omega_at_A+\omega_bt_B)}
E(\omega_a,\alpha)E(\omega_b,\beta)\,a_1^{\dagger}(\omega_a')
b_1^{\dagger}(\omega_b')\,;\ean $\ket{AB'}$ is obtained from
$\ket{AB}$ by replacing $\omega_b\leftrightarrow\omega_b'$ in $Z$,
or equivalently by relabelling the integration variables: \ban
\ket{AB'}&=& \int d\underline{\omega} d\underline{\omega}'\,
G(\omega_a,\omega_b')G(\omega_a',\omega_b)\,
Z(\omega_a,\omega_b,\omega_a',\omega_b')\ket{vac}\,.\ean
Obviously, by simply exchanging primed and unprimed integration
variables, $\ket{AB}=\ket{A'B'}$ and $\ket{AB'}=\ket{A'B}$, whence
the r.h.s. of (\ref{epsi4}). Inserting (\ref{epsi4}) into
(\ref{r4abstr}), we see that the quantity that we must compute is
\ba R_4(T_A,T_B)&=&I^2\,\int dt_Adt_B\,\big[\braket{AB}{AB}\,+\,
\big(\braket{AB}{AB'}+c.c.\big)\,+
\,\braket{AB'}{AB'}\big]\nonumber\\ &=&
R_{4,1}(T_A,T_B)\,+\,\Big[R_{4,2}(T_A,T_B)+c.c\Big]\,+ \,
R_{4,3}(T_A,T_B)\,. \label{r4start}\ea The integrals over $t_A$
and $t_B$ are exactly the same ones that we had in the calculation
of $R_2$.

The first term of the sum (\ref{r4start}) is the easiest one, and
can be given in closed form. In fact, in $\ket{AB}$ the integrals
over $\underline{\omega}$ and $\underline{\omega}'$ are factored,
and in addition, the integral on $\underline{\omega}$ gives the
same $c(t_A,t_B)$ that we met in the calculation of $R_2$, formula
(\ref{fctc}). That is, $\ket{AB}= c(t_A,t_B)\,\int
d\underline{\omega}'G(\omega_a',\omega_b')a_1^{\dagger}(\omega_a')
b_1^{\dagger}(\omega_b')\ket{vac}$. Consequently,
$R_{4,1}(T_A,T_B)\,=\,I\,R_2(T_A,T_B)\, \int d\underline{\omega}
\left|G(\omega_a,\omega_b)\right|^2$ where we recall that
$G(\omega_a,\omega_b)=g(\omega_a,\omega_b)
(1+e^{i(\omega_a+\omega_b)\tau})$. Anticipating over the
discussion of the next subsection, we use here the fact that the
terms that fluctuate in $\tau$ average to zero; so the last
integral is finally equal to $2\int d\underline{\omega}
\left|g(\omega_a,\omega_b)\right|^2=2J$. In conclusion, the first
term of the sum (\ref{r4start}) is \ba
R_{4,1}(T_A,T_B)&=&2\,I\,J\,R_2(T_A,T_B)\,. \ea The second term of
the sum (\ref{r4start}), $R_{4,2}(T_A,T_B)=I^2\int
dt_Adt_B\,\braket{AB}{AB'}$ gives \ba I^2\int d\underline{\omega}
d\underline{\omega}' d\underline{\tilde{\omega}}&&
f_A(\omega_a)f_A(\omega_a') f_B(\omega_b) f_B(\omega_b')\,
g^{*}(\tilde{\omega}_a,\tilde{\omega}_b)g^{*}(\omega_a',\omega_b')
g(\tilde{\omega}_a,\omega_b)g(\omega_a,\tilde{\omega}_b)
\nonumber\\ && \times\,
\left(1+e^{-i(\tilde{\omega}_a+\tilde{\omega}_b)\tau}\right)
\left(1+e^{-i(\omega_a'+\omega_b')\tau}\right)
\left(1+e^{i(\tilde{\omega}_a+\omega_b)\tau}\right)
\left(1+e^{i(\omega_a+\tilde{\omega}_b)\tau}\right)\nonumber\\
&&\times\,{\cal{E}}(\omega_a,\omega_b,\omega_a',\omega_b') \nonumber\\
&&\times
e^{-i(\omega_a-\omega_a')\tau}\,e^{-i(\omega_b-\omega_b')\tau}\,(\Delta
T)^2\,\mbox{sinc}[(\omega_a-\omega_a')\Delta T]
 \,\mbox{sinc}[(\omega_b-\omega_b')\Delta T]\,. \label{r42}\ea
The third term $R_{4,3}(T_A,T_B)=I^2\int
dt_Adt_B\,\braket{AB'}{AB'}$ gives \ba I^2\int d\underline{\omega}
d\underline{\omega}' d\underline{\tilde{\omega}}&&
f_A(\omega_a)f_A(\omega_a') f_B(\omega_b) f_B(\omega_b')\,
g^{*}(\tilde{\omega}_a,\omega_b')g^{*}(\omega_a',\tilde{\omega}_b)
g(\tilde{\omega}_a,\omega_b)g(\omega_a,\tilde{\omega}_b)
\nonumber\\ && \times\,
\left(1+e^{-i(\tilde{\omega}_a+\omega_b')\tau}\right)
\left(1+e^{-i(\omega_a'+\tilde{\omega}_b)\tau}\right)
\left(1+e^{i(\tilde{\omega}_a+\omega_b)\tau}\right)
\left(1+e^{i(\omega_a+\tilde{\omega}_b)\tau}\right)\nonumber\\
&&\times\,{\cal{E}}(\omega_a,\omega_b,\omega_a',\omega_b')  \nonumber\\
&&\times
e^{-i(\omega_a-\omega_a')\tau}\,e^{-i(\omega_b-\omega_b')\tau}\,(\Delta
T)^2\,\mbox{sinc}[(\omega_a-\omega_a')\Delta T]
\,\mbox{sinc}[(\omega_b-\omega_b')\Delta T]\,. \label{r43}\ea Note
that, as it should, the difference between $R_{4,2}(T_A,T_B)$ and
$R_{4,3}(T_A,T_B)$ is only in the contribution of the $G$'s, lines
one and two.

\subsection{Strategy of the calculation}

One cannot go beyond the formulae that we just derived for $R_2$
and $R_4$ without specifying what the evolution
${\cal{E}}(\omega_a,\omega_b,\omega_a',\omega_b')$ is (that is,
without specifying the setup) and without a simplification
strategy. Here is how this strategy goes \cite{hugues}.

\begin{enumerate}
\item We first notice that the times $T_A$, $T_B$ of interest are
typically $0$, $\tau$ etc; and as we said,
${\cal{E}}(\omega_a,\omega_b,\omega_a',\omega_b')$ is also a
product of terms containing either $1$ or some $e^{i\omega\tau}$.
So all our integrals for $R_2$ and $R_4$ are in fact sums of
integrals of the form $\int d\underline{\omega}\,{\cal
F}(\underline{\omega})\,e^{i\Omega(\underline{\omega})\tau}$.
Here, ${\cal F}$ is a product of $g(\omega,\omega')$'s (spectral
function of the pump and phase matching conditions) and of
cardinal sines $\mbox{sinc}((\omega-\omega')\Delta T)$ associated
to the time-resolution of the detectors; $\Omega$ is an algebraic
sum of the some of the integration variables $\underline{\omega}$.

\item Because we want the two pump pulses to be well-separated
(well-defined time-bins), it turns out that all the integrals in
which $\Omega\neq 0$ will average to zero. In fact, the typical
width of $g$ is $\frac{1}{t_c^{pump}}\geq \frac{1}{\Delta t}$,
where $t_c^{pump}$ and $\Delta t$ are, respectively, the coherence
time and the temporal width of each pump pulse $p(t)$. If the
time-bins are to be well-defined, we must impose $\tau>>\Delta t$.
Moreover, if one wants to distinguish the time-bins at detection,
one must also have a sufficiently small time-resolution for the
detector; so $\tau>>\Delta T$. In summary: in the frequency domain
(which is the integration domain), if $\Omega\neq 0$ the term
$e^{i\Omega(\underline{\omega})\tau}$ fluctuates with period
$\frac{1}{\tau}$, while in this range ${\cal
F}(\underline{\omega})$ is almost constant. The second step of the
calculation consists then in going through the factors to sort out
those integrals in which $\Omega=0$. This is the clever trick that
allows one to obtain readable formulae.

\item This being done, one can also perform the limit $\Delta
T\longrightarrow\infty$, leading to $\mbox{sinc}(x\Delta T)\simeq
\frac{1}{\Delta T}\delta(x)$. In fact, $x$ is of the form
$\omega-\omega'$, and this is in average close to the spectral
width of each down-converted photon $\frac{1}{t_c^{ph}}$. But a
detector cannot detect a photon unless $\Delta T>>t_c^{ph}$. This
is the precise meaning of the formal limit $\Delta
T\longrightarrow\infty$. Obviously this limit must be performed
{\em after} the estimate described in point 2.
\end{enumerate}
In summary, for each of the setups that we want to study, we must
replace ${\cal{E}}(\omega_a,\omega_b,\omega_a',\omega_b')$ with
its explicit value, then by inspection identify those terms for
which the dependance in $\tau$ identically vanishes under the
integral.

\subsection{Calculations for section \ref{seccal}}

For the calibration setup of section \ref{seccal}, we must compute
$R(0,0)$ and $R(\tau,0)$. Here,
${\cal{E}}(\omega_a,\omega_b,\omega_a',\omega_b')=1$ because the
modes don't evolve from the preparation to the detection. Then,
$R_2$ given in (\ref{r2tout}) is the sum of four integrals because
of the product in line two; while $R_{4,2}$ and $R_{4,3}$ given
respectively by (\ref{r42}) and (\ref{r43}) are the sum of sixteen
integrals because of the products in line two.

Let's set $T_A=T_B=0$, and look first at $R_2$. Only the product
$1\times 1$ gives an integral whose argument does not contain
$\tau$, so we forget about the three other integrals. Through the
limit $\mbox{sinc}(x\Delta T)\simeq \frac{1}{\Delta T}\delta(x)$,
we obtain $\omega_j=\omega_j'$ and consequently
$R_2(0,0)=I\,J_{AB}$. So we have obtained
$R(0,0)=I\,J_{AB}+O(I^2)$ as announced.

If we set $T_A=\tau$, $T_B=0$, it is easy to become convinced that
none of the four integrals that compose $R_2$ can become
independent of $\tau$; therefore, $R_2(\tau,0)=0$ and we must
compute $R(\tau,0)=R_4(\tau,0)$. Obviously, $R_{4,1}(\tau,0)=0$
because it is proportional to $R_2$. By inspection, one sees that
$R_{4,2}(\tau,0)=0$ as well since none of the sixteen integrals
can be made independent of $\tau$. In $R_{4,3}(\tau,0)$, only the
integral associated to the product $1e1e$ (with obvious notations)
is independent of $\tau$. For this integral, we can again set
$\omega_j=\omega_j'$ and we obtain finally
$R(\tau,0)=R_{4,3}(\tau,0)=I^2J_AJ_B$ as announced.

\subsection{Calculation for section \ref{secfran}}

For the Franson interferometer of section \ref{seccal}, we must
compute $R(\tau,\tau)$. Here however, \ba
{\cal{E}}(\omega_a,\omega_b,\omega_a',\omega_b')&=&\left(1-e^{i(\omega_a\tau+\alpha)}\right)
\left(1-e^{i(\omega_b\tau+\beta)}\right)
\left(1-e^{-i(\omega_a'\tau+\alpha)}\right)
\left(1-e^{-i(\omega_b'\tau+\beta)}\right)\,,\ea where we dropped
a global factor $\frac{1}{2^4}$. Consequently, $R_2$ given in
(\ref{r2tout}) is the sum of $2^6=64$ integrals, while $R_{4,2}$
and $R_{4,3}$ given respectively by (\ref{r42}) and (\ref{r43})
are the sum of $2^8=256$ integrals.

Let's look at $R_2$. By inspection, one finds that the integrals
whose argument is independent of $\tau$ are four: $ee|1111$, that
gives a contribution $1$; $11|eeee$, that also gives a
contribution $1$; $1e|ee11$, whose contribution is
$e^{i(\alpha+\beta)}$; and $e1|11ee$, whose contribution is
$e^{-i(\alpha+\beta)}$. In these notations, the first two items
correspond to the products of terms of line two, the last four
items correspond to the products within ${\cal{E}}$. Finally,
performing the limit $\Delta T\longrightarrow \infty$ we find \ba
R_2(\tau,\tau)&=& 2\,I\,J_{AB}\, \big(1+\cos(\alpha+\beta)\big)\ea
that is indeed (\ref{r2fin}) up to a multiplicative factor $2$.
Immediately then we have also \ba R_{4,1}(\tau,\tau)&=&
4\,I^2\,J_{AB}\, \big(1+\cos(\alpha+\beta)\big)\ea accounting for
the first term of the r.h.s. of (\ref{r4fin}).

Moving to $R_{4,2}$, by inspection, one can verify that the only
integrals that will not average to zero are those associated to
the following four products: $1111|eeee$ and $eeee|1111$, both
giving 1; $1e11|ee11$, that gives $e^{i(\alpha+\beta)}$; and
$e1ee|11ee$, that gives $e^{-i(\alpha+\beta)}$. As before, in
these notations the first four items represent the terms of line
two of (\ref{r42}), the last four items correspond to the products
within ${\cal{E}}$. After the usual limit, one finds \ba
R_{4,2}(\tau,\tau)+c.c&=&
2\,I^2\,J_4\,\big(1+\cos(\alpha+\beta)\big)\ea accounting for the
second term on the r.h.s. of (\ref{r4fin}).

As for $R_{4,3}$, again only four integrals out of 256 will not
average to zero, namely those associated to $1111|eeee$,
$eeee|1111$, $1e1e|e1e1$ and $e1e1|1e1e$; however here, all these
contributions give simply $1$, so finally \ba
R_{4,3}(\tau,\tau)&=& 4\,\,I^2\,J_AJ_B \ea that is the last term
in the r.h.s. of (\ref{r4fin}). This concludes our demonstration.

\end{appendix}


\begin{thebibliography}{99}
\bibitem{noteper} In Sections \ref{secgen}-\ref{secexp} of this paper,
where we focus on time-bin entanglement, we call "pulse" each
pulse that enters the non-linear crystal; therefore, one qubit is
created by two pulses. In Ref. \cite{ivanpra}, we had used "pulse"
to describe each pulse produced by the pump laser, each such
process creating a qubit. So "per pulse" in Ref. \cite{ivanpra} is
equivalent to "per qubit" here.
\bibitem{note0} This efficiency factor takes into account the
losses on the lines, the quantum efficiency $\eta$ of the
detector, and the amplitude of each monochromatic field component
(whose weak $\nu$-dependance has been omitted).
\bibitem{notenew} Rigorously, $\theta=\varepsilon\,\omega$. However,
since we have extracted the main contribution $\omega\tau$,
we have $\varepsilon\,\Delta\omega<<\pi$, with $\Delta\omega$
the spectral width of the photons ($\Delta_{A,B}$). Consequently,
we set $\theta=\varepsilon\,\bar{\omega}$ with $\bar{\omega}$
the central frequency of the photon.
\bibitem{note3} Again, we give coincidence rates up to global
multiplicative factors. In particular, the multiplicative factor
omitted in Section \ref{seccal} is not the same as the one omitted
here: as can be found in the appendix, to compare the rates in the
two Sections, the rates presented in this Section must be
multiplied by $\frac{1}{8}$. This factor is indeed what one
expects, because suppose that $R_c$ is the rate of detection of a
pair in the setup of Section \ref{seccal}. For the Franson
interferometer, in only $\frac{1}{4}$ of the cases both photons
will arrive in the intermediate time-bin, and there are four
possibly detection patterns; but the energy available to create a
pair is the energy of two pulses; so indeed the detection rate
(averaged over all choices of $\alpha+\beta$) in the Franson
setup, supposing the same losses in the lines, is
$R_F\,=\,\big(\frac{1}{4}\times \frac{1}{4}\times 2\big)R_c =
\frac{1}{8}R_c$.
\bibitem{note1} The data obtained in the early experiment (full
squares in Fig. \ref{results}) are more noisy than the new ones
(open circles) because the crystal that we used in 2001 was less
efficient and because in the meantime we have improved the
techniques to stabilize the interferometers.
\end{thebibliography}
\end{document}